\documentclass[twocolumn,epsfig,pre]{revtex4}

\usepackage{graphics}
\usepackage{graphicx}
\usepackage{epstopdf}
\usepackage{amssymb}
\usepackage{amsmath}
\usepackage{hyperref}
\usepackage{latexsym}
\usepackage{color}

\begin{document}

\renewcommand{\thefootnote}{\fnsymbol{footnote}}
\renewcommand{\theequation}{\arabic{section}.\arabic{equation}}

\title{Energy polydisperse 2$d$ Lennard-Jones fluid in presence of flow field}

\author{Lenin S.~Shagolsem}
\email{slenin2001@gmail.com}
\affiliation{Department of Physics, National Institute of Technology Manipur, Imphal, India} 

\date{\today}

\begin{abstract}

\noindent {The behavior of energy polydisperse $2d$ Lennard-Jones fluid (in thin-film geometry) is studied subjected to linear flow field using molecular dynamics simulations. By considering neutral and selective substrates we systematically explore the effect of flow field on particle ordering as well as response of the system. It is shown that particle density profile, spatial organization as well as local particle identity ordering in the film are affected. Furthermore, we observe flow field induced melting associated with a decrease of effective interaction parameter, $\left< \epsilon_i^{\rm eff} \right>$, which characterizes local neighborhood identity ordering. In terms of macroscopic response, the systems show both shear-thinning and shear-thickening behaviors, and shear-thinning exponent decreases with increasing temperature and eventually attains Netwonian fluid-like behavior at sufficiently high temperature. It is found that the qualitative behaviour of one component LJ-fluid and energy polydisperse fluid with neutral substrates are similar in many respects, while the one with selective substrate shows differences. In the case of energy polydisperse system, the effect of having different substrate types is significantly manifested in the density profile near the interface, $\left< \epsilon_i^{\rm eff} \right>$, and in the viscosity. We have shown that, unlike one component fluid, it is possible to tune the macroscopic response by tuning substrate-fluid interaction in energy polydisperse fluids.} 
\end{abstract}

\maketitle


\section{Introduction}
\label{sec: introduction}

Multi-component fluids are very common in nature, where degree of polydispersity is often attributed to a distribution of particle sizes. The outcome of the particle size variation has an important role, e.g., in the glass forming ability of the colloidal fluids. In experiments, polydispersity offers a means to avoid crystallization and thus allows one to study very high viscous fluids.\cite{pusey1987,auer2001} Application of these materials demands understanding their physics and our ability to manipulate them for industrial processing. Various equilibrium aspects of size polydisperse systems (e.g., packing, thermodynamics, etc.) have been discussed in earlier works.\cite {Stell, Sollich, Warren,Salacuse1982_JCP77,Salacuse1984_JCP81,Sollich1998_PRL80,Warren1998_PRL80,Evans2001_JCP114,Sollich2002_JPCM14, Wilding2006_JCP125,Buzzacchi2006_PRE73,Wilding2008_PRE77,Jacobs2013_JCP139}  On the other hand, polydispersity is very common in biological systems as well, where a cell contains large amount of different types of macromolecules that differ in their shape, size, and interactions leading to interesting dynamical behavior and geometry controlled kinetics.\cite{IMSokolov2012_SoftMatter,RMetzler2000_PhysicsReports,eli2012_PhysToday,IIzeddin2014_eLife,Benichou,jeon2016} In our recent work,\cite{Shagolsem2015_JCP,Shagolsem2016_JCP} a model complex fluid characteristic of biological systems was introduced, where we focused on interaction multiplicity (i.e., energy polydispersed) among the constituents -- a minimal physical model in which {\it all particles are different} (APD) in the sense that their interaction parameters are chosen at random from a given distribution. The results of our earlier work on statics and dynamics of APD fluid close to solidification transition are briefly discussed below. 

In an energy polydisperse fluid, although the particles have random interactions, the local self-organization show that the particles relax into a non-random state characterized by clustering of particles according to the values of their pair interaction parameters which become more pronounced as temperature is decreased throughout the fluid region. This particle identity ordering has consequences in the particles' dynamical behavior. For example, when the particle identity ordering sets in, different types of particles find themselves in different local environments and thus show distinct diffusion coefficients. It is important to note that size polydispersity leads to entropic effects associated with packing a large number of dissimilar objects in an efficient way thus increasing the number of available configurations, whereas interaction polydispersity affects the weights of the different configurations through neighborhood identity ordering. Furthermore, several aspects of the system, viz., identity ordering and metastable cluster, and effect on tensile strength (where a multicomponent model solids shows limiting tensile strength that far exceed the corresponding one-component solids) are reported in references~\cite{azizi2019,kulveer2019}. 

In the present paper, non-equilibrium aspects of the energy polydispersed fluid in thin-film geometry is addressed, where we study energy polydisperse Lennard-Jones fluid subjected to linear flow field. There is large volume of work on the rheological studies of colloidal dispersion and soft-matter in general.\cite{brown2014,chen2010} Recent advances in the field allow access to the microscopic structural changes during the process and thus gives better understanding of the mechanical response of these complex fluids. In particular, we study the effect of flow field on particle distribution in the film, spatial ordering of particles, and local identity ordering.  Also, macroscopic response is studied by measuring stress components and hence shear viscosity. Finally, the role of substrate quality on the aforesaid quantities of the fluid is also investigated by considering two types of substrates (non-selective and selective substrates) described in the following section. \\

Our paper is organized as follows: In section~\ref{sec: model-simulation-details}, we present the model and simulation details. Particle distribution in the film through density profile is reported in section~\ref{sec: density-profile}, particle ordering and flow-field induced melting in section~\ref{sec: identity-ordering}, and macroscopic response in section~\ref{sec: macro-response}. Finally, in section~\ref{sec: conclusion} we summarize and discuss the results of this work. 


\section{Model and simulation details}
\label{sec: model-simulation-details}

We perform coarse-grained MD simulations (NVT-ensemble) in two dimensions considering $N=2500$ particles in a simulation box of dimension $L_x=80\sigma$ and $L_y=49\sigma$, with $\sigma$ particle diameter. All the particles have same mass $m$ and size $\sigma$ which are set to unity.  
The particles interact via Lennard-Jones (LJ) potential 
\begin{equation}
U_{ij}(r) = 4\epsilon_{ij}\left[(\sigma/r)^{12}-(\sigma/r)^{6}\right]~,
\label{eqn: LJ-potential}
\end{equation}
which is cut-off and shifted to zero at $r_{\rm c}=2.5\sigma$. Here, $\epsilon_{ij}$ and $r$ are the interaction strength and separation between a pair of particles $i$ and $j$ respectively. 
The equation of motion of $i^{\rm th}$ particle is given by the Langevin equation
\begin{equation}
 m \frac{d^2 \bf r_i}{dt^2} + \zeta \frac{d \bf r_i}{dt} = -\frac{\partial U}{\partial \bf r_i} + {\bf f_i}~,
 \label{eqn: langevin}
\end{equation}
with $\bf r_i$ the position of particle $i$, $\zeta$ the friction coefficient (assumed to be the same for all particles), $U$ is the sum of all the pair potentials $U_{ij}$ ($i\ne j$), and $\bf f_i$ a random external force with zero mean and second moment proportional to the product of temperature $T$ and $\zeta$. All the physical quantities are expressed in LJ reduced units, with LJ time $\tau_{_{\rm LJ}}=1$.\cite{allen} We use velocity-Verlet scheme to integrate the equations of motion with time step $\delta t = 0.005$. \\

\begin{figure}[ht]
	\includegraphics*[width=0.45\textwidth]{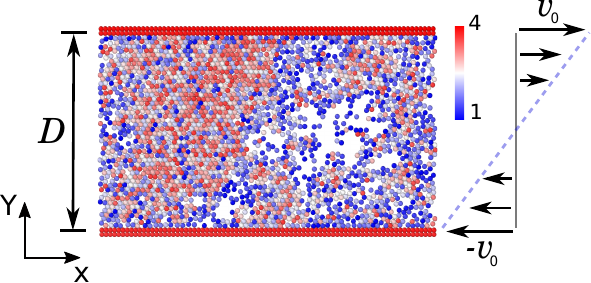} 
	\begin{center}
		\caption{Equilibrium configurations of the APD system relaxed at $\delta=0$ for neutral substrate. Fluid flows along X-direction and the velocity gradient is along the Y-direction. Wall moves along with the fluid. Particles are colored according to their interaction parameter $\epsilon_i$ in the range 1-4, see color scale bar.} 
		\label{fig: fig1_schematic} 
	\end{center}
\end{figure}

In order to introduce energy polydispersity, we assign interaction parameter $\epsilon_i$ to each particle $i$ which is drawn randomly from a uniform distribution in the range 1-4, where the value of interaction strength $\epsilon_i$ defines the particle identity. Between particles $i$ and $j$ the interaction strength is given by $\epsilon_{ij}=\sqrt{\epsilon_i\epsilon_j}$ (following Lorentz-Berthelot anzatz \cite {Berthelot}). The distribution $P(\epsilon_{ij})$ is shown in fig.~\ref{fig: fig1_schematic}(a), where the most probable value $\epsilon_{ij}^{\rm max}=2.0$ and mean $\overline{\epsilon}_{ij}=2.42$. As a reference system, we simulate a one-component (1C) LJ system at the same density with interaction parameter $\epsilon_{ij}=2.5$ (mid-point of the interval $1-4$). As reported in earlier studies, \cite{Shagolsem2015_JCP,Shagolsem2016_JCP} at this number density $\rho^\ast=0.69$ the system behaves as a liquid at high temperatures. In general, the phase diagram of APD system differs from that of a pure one component (1C) fluid, \cite{Wilding2008_PRE77} where in 1C system (at the same density) it passes through the triple point and thus goes directly from fluid to solid-gas coexistence region upon cooling.\cite{Abraham1981_PhysRep,patashinsky} 
The liquid-solid transition takes place within a narrow temperature range in which the mean interaction energy decreases sharply upon cooling from the liquid to solid-fluid coexistence region. The transition temeperature is estimated to be $T^\ast \approx 1.0$ for both APD and 1C systems (for bulk systems), and interesting dynamical differences are observed close to $T^*$ as detailed in references \cite{Shagolsem2015_JCP,Shagolsem2016_JCP}. \ 

\begin{table}[h]
	\caption{Pair-wise interaction strength among the species}
	\centering
	\begin{tabular}{|l|l|l|}
		\hline
		\bf{Between $i-j$} & \bf{Strength $\epsilon_{ij}$} & \bf{Nature} \\
		\hline
		fluid - fluid & polydisperse & attractive (different strength) \\
		\hline
		fluid - wall & (a) $2.5$ & attractive (same strength) \\
		             &       & -- {\it non-selective wall case}  \\ 
		             & (b) polydisperse & attractive (different strength) \\
		             &       & -- {\it selective wall case} \\
		\hline
		wall - wall & $1.0$ & repulsive \\
		\hline
	\end{tabular}
	\label{table: params}
\end{table}

In the current study, fluid particles are confined by two all-atom walls in slit geometry of width $D=46\sigma$, see fig.~\ref{fig: fig1_schematic}, and thus the system is periodic along X-axis only. Two types of confining walls are considered: (i) {\it non-selective walls} (ns-Wall), where the wall particles have no preferential interaction with the fluid particles, and we set wall-fluid interaction strength $\epsilon_{\rm wall - fluid} = 2.5$ for all particle types, and (ii) {\it selective walls} (s-Wall), where the wall-fluid interaction strength varies depending on the particle type. Among the wall particles we set $\epsilon_{\rm wall}=1.0$. In both the wall types, interaction cut-off distance is set to $r_c=2.5\sigma$. However, in this study, we do not consider the case of ns-W made by setting $r_c=1.12\sigma$ (i.e., purely repulsive case). Various interaction strengths prensent in the system are briefly summarized in table~\ref{table: params}.

It is known that the presence of substrate can supress or enhance onset of solidification and thus shift the transition temperature to a lower or higher values. 
For example, simulations study of freezing and melting of Lennard-Jones methane confined within slit pores by Miyahara and Gubbins show that when the adsorbate-wall interactions were weaker compared to the adsorbate-adsorbate interactions, the freezing temperature is significantly lower than in the bulk, as reported in most of the experiments for silica pores. In contrast, a large increase in the freezing temperature is observed for strongly attractive materials, such as porous-carbons. No appreciable change in the freezing temperature is observed when the adsorbate-wall and adsorbate-adsorbate interactions have similar strength. 
A detailed investigation is necessary to understand the effect of confinement on the solidification transition of APD fluids under variation of parameters such as density, film thickness, etc., and we leave this task for a future study. However, for the purpose of current work, approximate values of $T^*$ is located for the confined systems by performing heating/cooling cycles and using the procedure detailed in reference \cite{Shagolsem2015_JCP}. 
For the considered density, it is observed that $T^*$ deviates from that of the bulk (albeit very small) and follow the trend: $T^*_{\rm bulk}(\approx 0.98) < T^*_{\rm s-w}(\approx 0.997) < T^*_{\rm ns-w}(\approx 1.01)$. Since the walls are attractive elevation of $T^*$ compared to the bulk is expected. However, as defined above, ns-Wall attracts all the particles with same strength and s-Wall, depending on particle type, attracts with different strengths. Consequently, particle enrichment near the walls is more for ns-Wall type and thus expect to solidify at a higher temperature, and for stronger polymer-wall interaction strength $T^*$ is expected to increase further. Since our interest is in the fluid phase we focus our attention to $T\ge T^*$ (otherwise stated clearly). \\

In order to study the effect of flow field, we introduce a linear velocity profile defined as 
\begin{equation}
v(y) = \frac{2 v_0}{D}\left(y-\frac{D}{2}\right)~,
\end{equation}
where $y$ varies from $0$ to $D$, as shown in figure~\ref{fig: fig1_schematic}.  Furthermore, the upper(lower) walls are moving with velocity $+v_0(-v_0)$, and the shear rate is defined as 
\begin{equation}
\dot{\gamma}=\frac{2v_0}{D}~.
\end{equation}
It is important to note that by imposing linear flow profile and moving the walls simultaneously we neglect the slipping near walls. 
Typically, we relax the sample for $2.5\times 10^4\tau_{_{\rm LJ}}$ followed by shear for $5\times 10^5\tau_{_{\rm LJ}}$ corresponding to 100 million MD time steps. The results reported here are obtained in the steady state. The simulations are carried out using open source program LAMMPS.\cite{lammps}



\section{Density profile}
\label{sec: density-profile}

\begin{figure}[ht]
	\includegraphics*[width=0.4\textwidth]{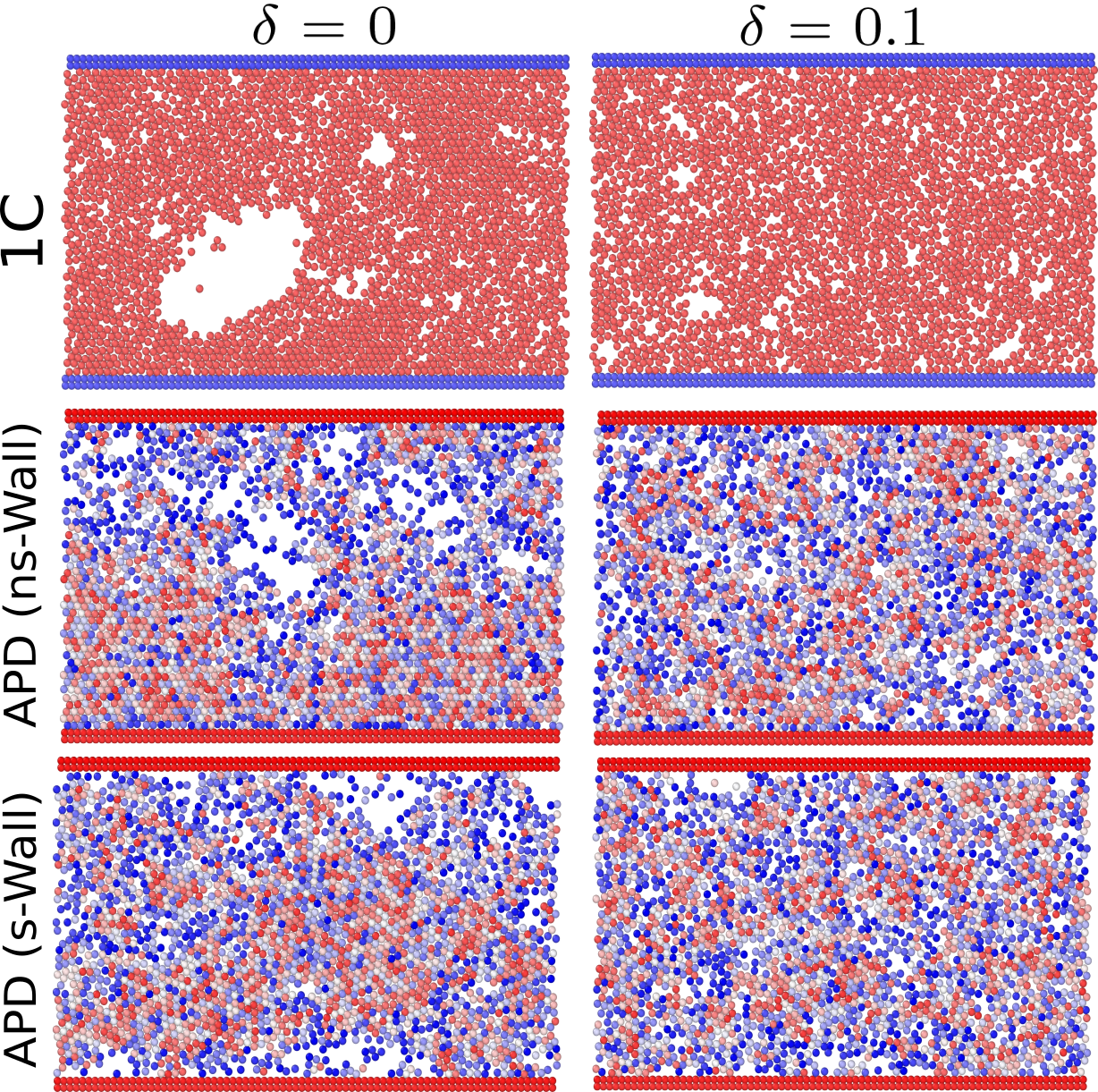}
	\begin{center}
		\caption{Typical configurations of the 1C and APD systems in the steady state for shear rate $\dot{\gamma}=0.003$ shown for (A) $\delta=-0.1$, (B) $\delta=0$, and (C) $\delta=0.1$. Particles are colored according to their $\epsilon_i$ values, with $\epsilon_i=1$(blue) and $\epsilon_i=4$(red), see color scale bar.} 
		\label{fig: simu-snapshot} 
	\end{center}
\end{figure}

Typical steady state configurations of the system for $\dot{\gamma}=0.003$ at temperatures below, at, and above the transition, i.e., $\delta\equiv \frac{T-T^*}{T^*}=-0.1,0,0.1$, respectively are shown in figure~\ref{fig: simu-snapshot}. For $T>T^*$ the particles are homogeneously disributed in the film and with decrease in temperature the system becomes inhomogeneous (i.e., development of voids) and the particles with large $\epsilon_i$ values clumped together forming a core and smaller $\epsilon_i$ particles surrounding it. As observed in the equilibrium studies,\cite{Shagolsem2015_JCP} this particle identity ordering is expected at small and moderate shear rates. However, it is interesting to note that the difference in substrate quality has effect on the particle distribution near the interface. 

\begin{figure}[ht]
 	\includegraphics*[width=0.45\textwidth]{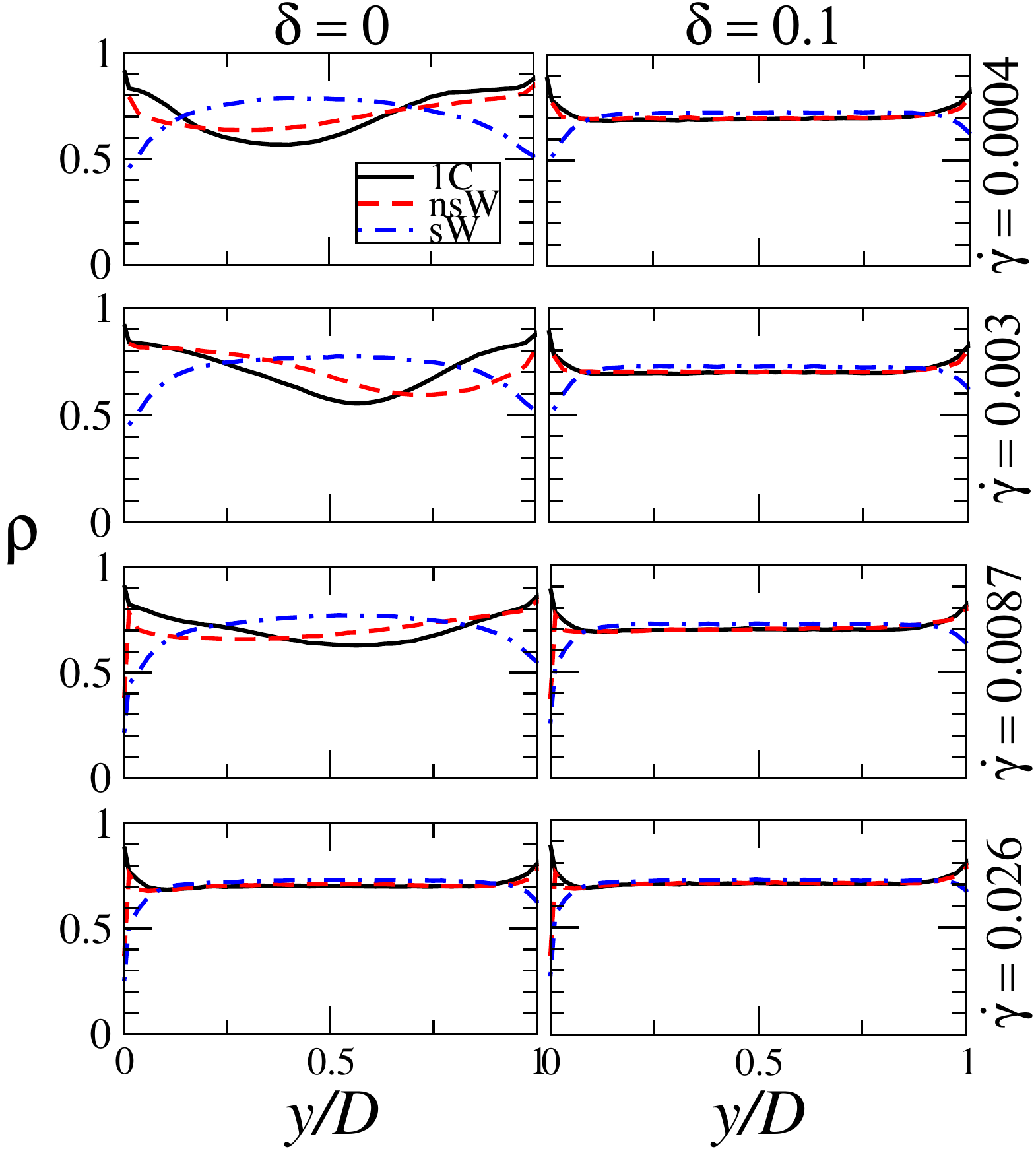}
 	\begin{center}
 		\caption{Laterally averaged particle density $\rho$ at different positions in the film indicated by the normalize distance $y/D$. The walls are located at $y=0$ and $y=1$.} 
 		\label{fig: apd-rho-compare} 
 	\end{center}
 \end{figure}

To understand the particle distribution in the film we calculate the averaged density profile at different positions of the film and compare with 1C system. In order to obtain the density $\rho$ as a function of distance $y$ from the walls, we divide the film into several layers and calculate the density in each layer. In figure~\ref{fig: apd-rho-compare}, we compare the density profiles of 1C and APD systems obtained at different values of shear rates and three different values of temperatures $\delta =-0.1,0,0.1$. 

As expected, at high temperature the particle distribution is homogeneous and hence $\rho(y)$ is flat in the bulk, see figure~\ref{fig: apd-rho-compare} for $\delta=0.1$. However, in the wall-particle interface there is particle enrichment for both 1C and APD with non-selective walls, as shonw by the slight increase in $\rho(y)$ as $y$ approaches walls. Whereas decrease of $\rho(y)$ near walls is seen for APD with selective substrate indicating particle depletion. No shear rate dependence on the density profile is observed at this temperature. \\

However, at lower temperatures ($\delta=0,-0.1$) the difference in the profile is enhanced both in bulk and interface regions. For 1C and APD (ns-Wall) the density profile develops a minimum and at further lower temperature for APD (ns-Wall) and relatively high shear rates particles there is strong depletion near interface. It is interesting to note that for APD (s-Wall) the density profile remains qualitatively the same, i.e. maximum in the bulk and minimum at the interfaces, and weak dependence on the shear rate. In the following, we discuss the spatial ordering of particles and local identity ordering. 

\section{Particle ordering and shear induced melting}  
\label{sec: identity-ordering} 

To characterize the spatial organization of particles we calculate the radial distribution function defined as
\begin{equation}
g(r)=\frac{1}{\rho_{_0} N}\left<\sum_{i}\sum_{j\ne i} \delta(r-r_{ij}) \right>~,
\end{equation}
where $\rho_{_0}$ is the particle number density in the system, and $r_{ij}=r_i-r_j$ distance between $i^{th}$ and $j^{th}$ particles.  
\begin{figure}[ht]
	\includegraphics*[width=0.45\textwidth]{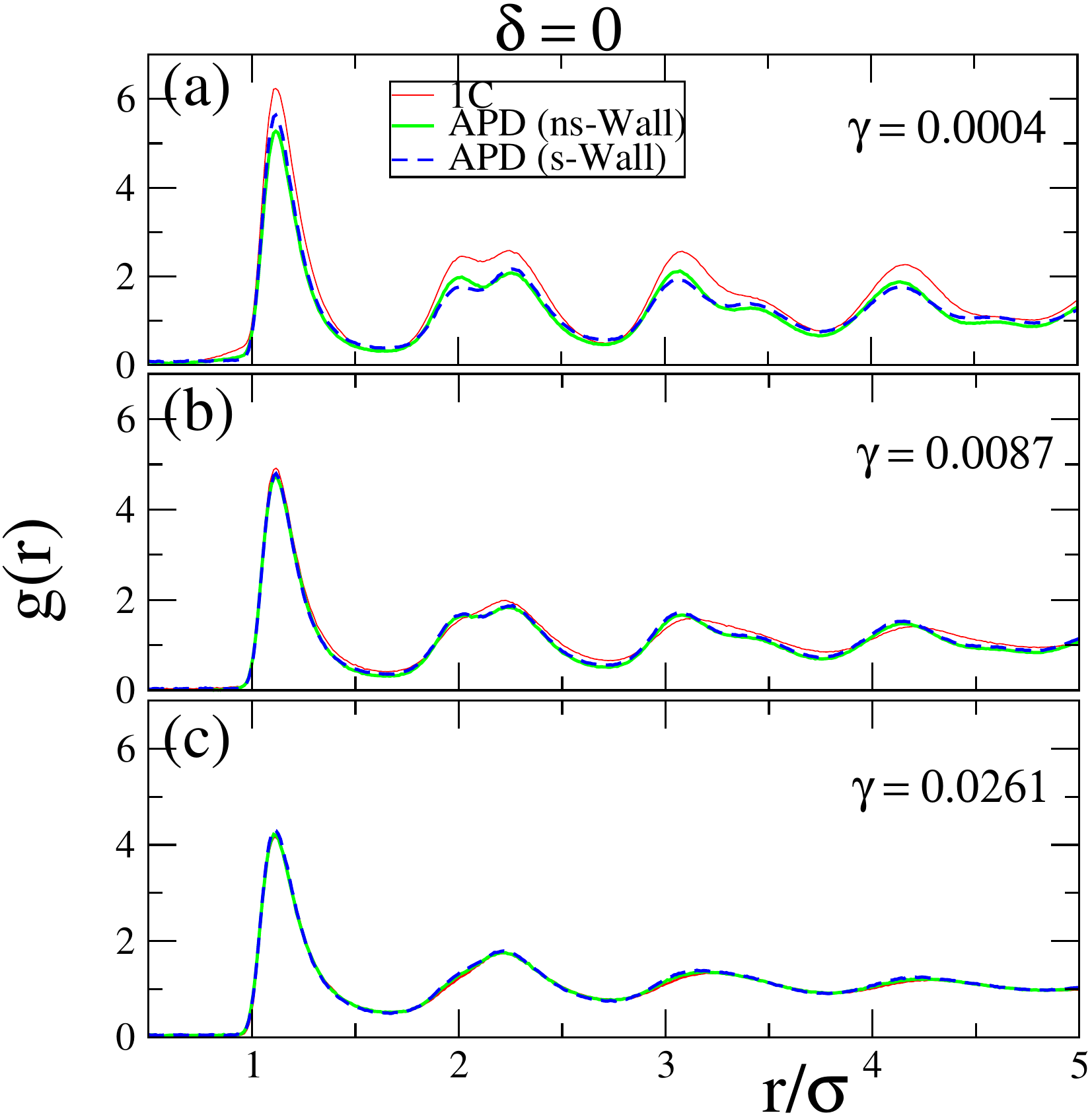} 
	\begin{center} 
		\caption{Radial distribution function $g(r)$ for (a) 1C system, (b) APD with ns-Wall, and (c) APD with s-Wall obtained at $\delta=-0.1$ for different values of shear rates $\dot\gamma$ indicated in the figure. In all the systems, $g(r)$ becomes liquid-like at high shear rate indicating shear-induced melting.}
		\label{fig: gr2d} 
	\end{center} 
\end{figure}

Since for $T>T^*$ or $\delta > 0$ the system is in liquid state we focus on $T<T^*$ (where the system is in solid-gas coexistence) in order to understand the effect of shear on spatial organization. In figure~\ref{fig: gr2d}, $g(r)$ is displayed for 1C and APD systems obtained at $\delta=-0.1$ for different values of shear rates. Higher order peaks corresponding to hexagonal packing are observed for small $\dot\gamma$ values which is gradually transformed to liquid-like ordering upon increasing $\dot\gamma$. Such shear-induced melting and distortion of $g(r)$ when subjected to shear-flow in simple liquids were observed in earlier works.\cite{hess1980,evans1984,hess1985} The associated effects in the transport coefficients due to structural reorganization is discussed in section~\ref{sec: macro-response}. \\ 

Another aspect of particle ordering is the identity of neighboring particles. Neighborhood identity ordering is quantified by means of effective interaction parameter of particle $i$, defined as
\begin{equation}
\epsilon_i^{\rm eff}=\frac {1}{n_b}\sum_{j=1}^{n_b}\epsilon_{ij}~,
\label{eqn: epsilon}
\end{equation}
where the sum over $j$ goes over all the $n_b$ neighboring particles within a cut-off radius $r_c=1.7$ which roughly corresponds to the minimum between the first and the second peaks of the radial distribution function. The value of $\epsilon_i^{\rm eff}$ ranges from 0 (for no neighbors within $r_c$) to 4. \\
\begin{figure}[ht]
\includegraphics*[width=0.5\textwidth]{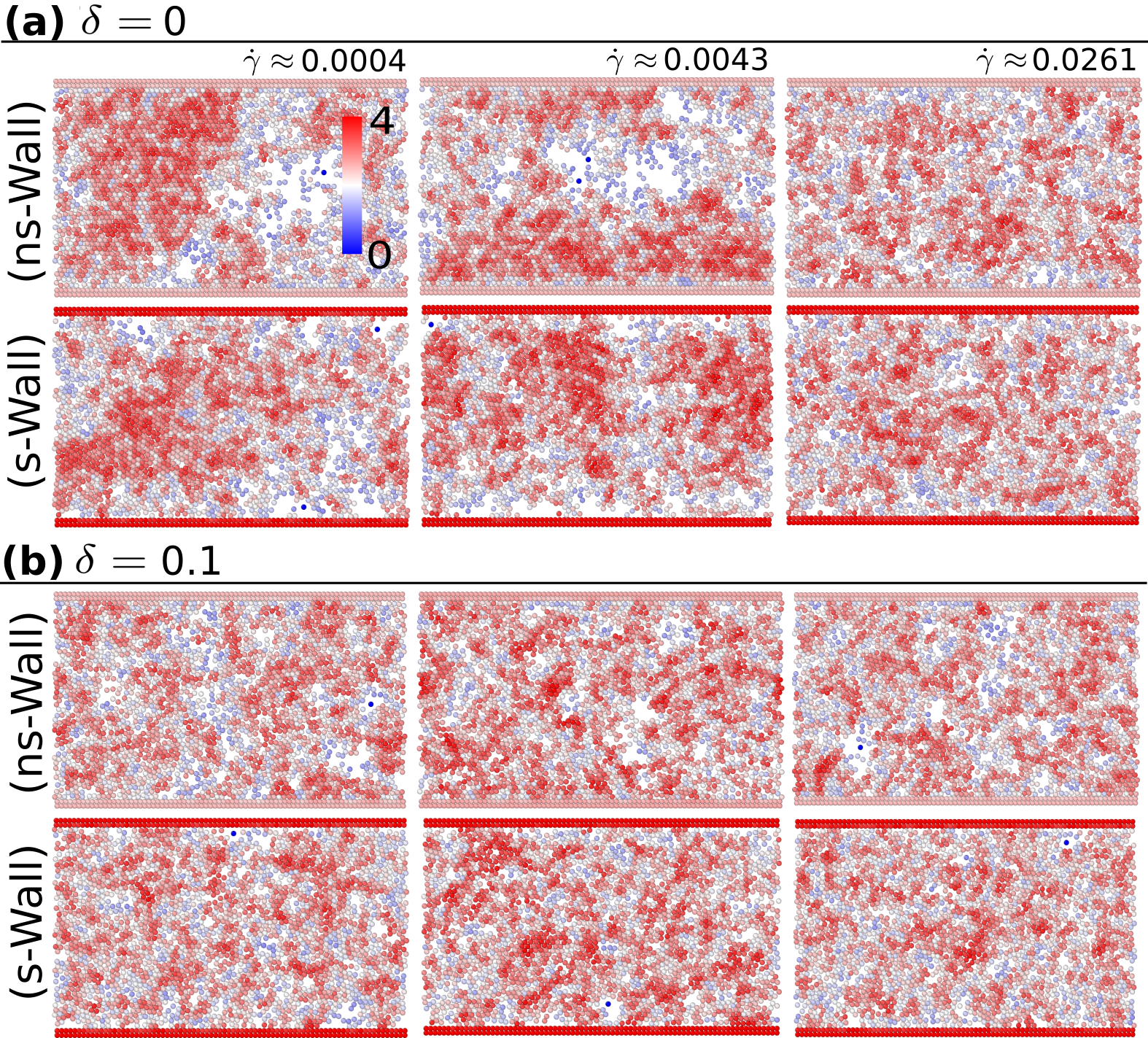}
\begin{center} 
\caption{Typical steady state configurations of APD systems shown for three different values of shear rates $\dot{\gamma}\approx 0.0004,~0.0043,~0.0261$ at (a) $\delta=0$ and (b) $\delta=0.1$. Particles are colored according to their effective interaction parameter $\epsilon_i^{\rm eff}$ values. See color scale bar for $\epsilon_i^{\rm eff}$ values.} 
\label{fig: configs-under-shear} 
\end{center} 
\end{figure} 

In figure~\ref{fig: configs-under-shear}, we display the systems under shear, where particles are colored according to their $\epsilon_i^{\rm eff}$, shown for different shear rates at $\delta=0,~0.1$. One can see in the figure, for both selective and non-selective walls cases, there are regions of high $\epsilon_i^{\rm eff}$ surrounded by regions with low $\epsilon_i^{\rm eff}$ values. For $\delta>0$ particles are homogeneously distributed in the film (as discussed above) and observe smaller regions of relatively high $\epsilon_i^{\rm eff}$ values, and as $T\rightarrow T^*$ high $\epsilon_i^{\rm eff}$ regions gets bigger. 

\begin{figure}[ht]
	\includegraphics*[width=0.45\textwidth]{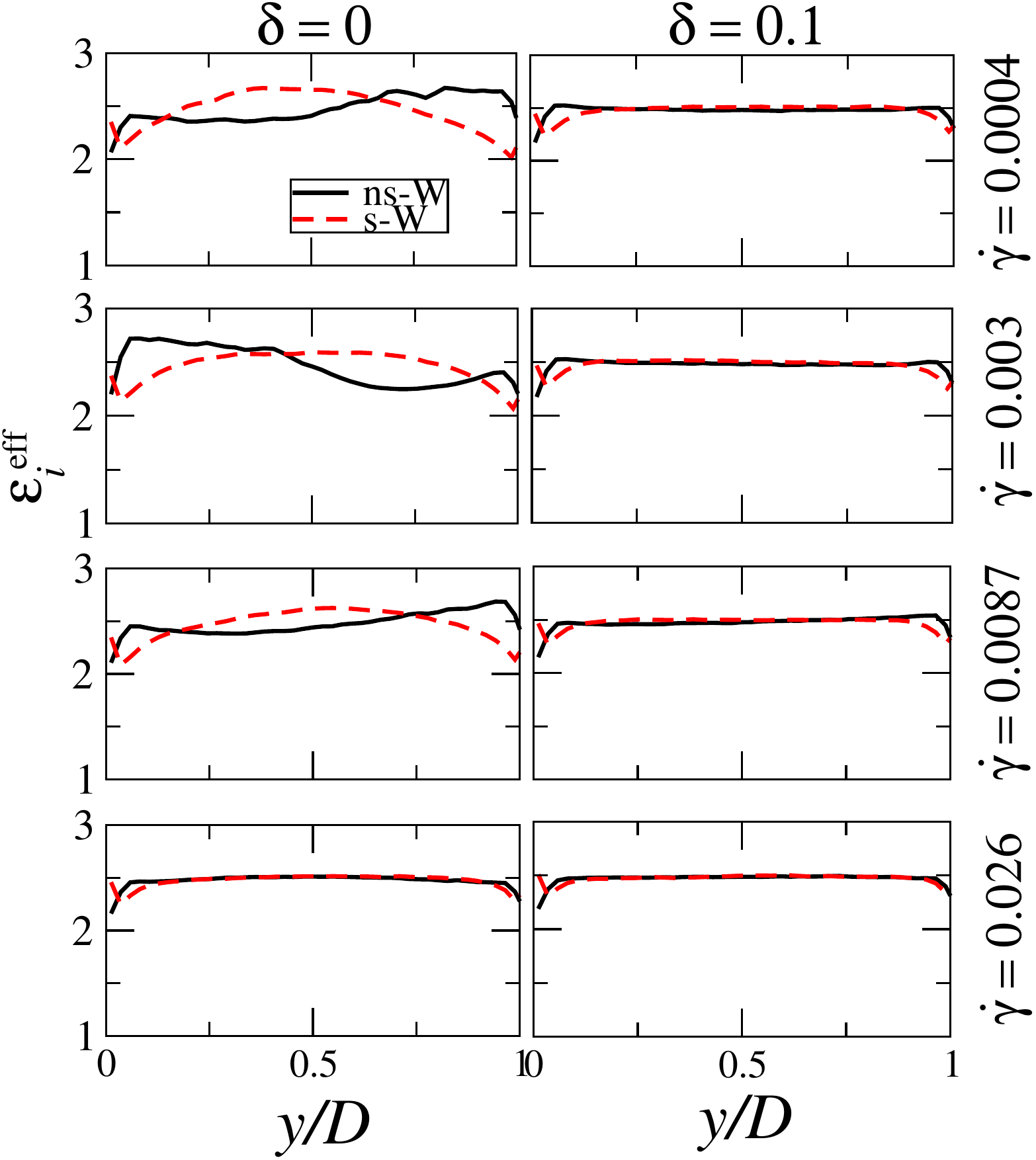} \\ \vspace{0.2cm}
	\includegraphics*[width=0.45\textwidth]{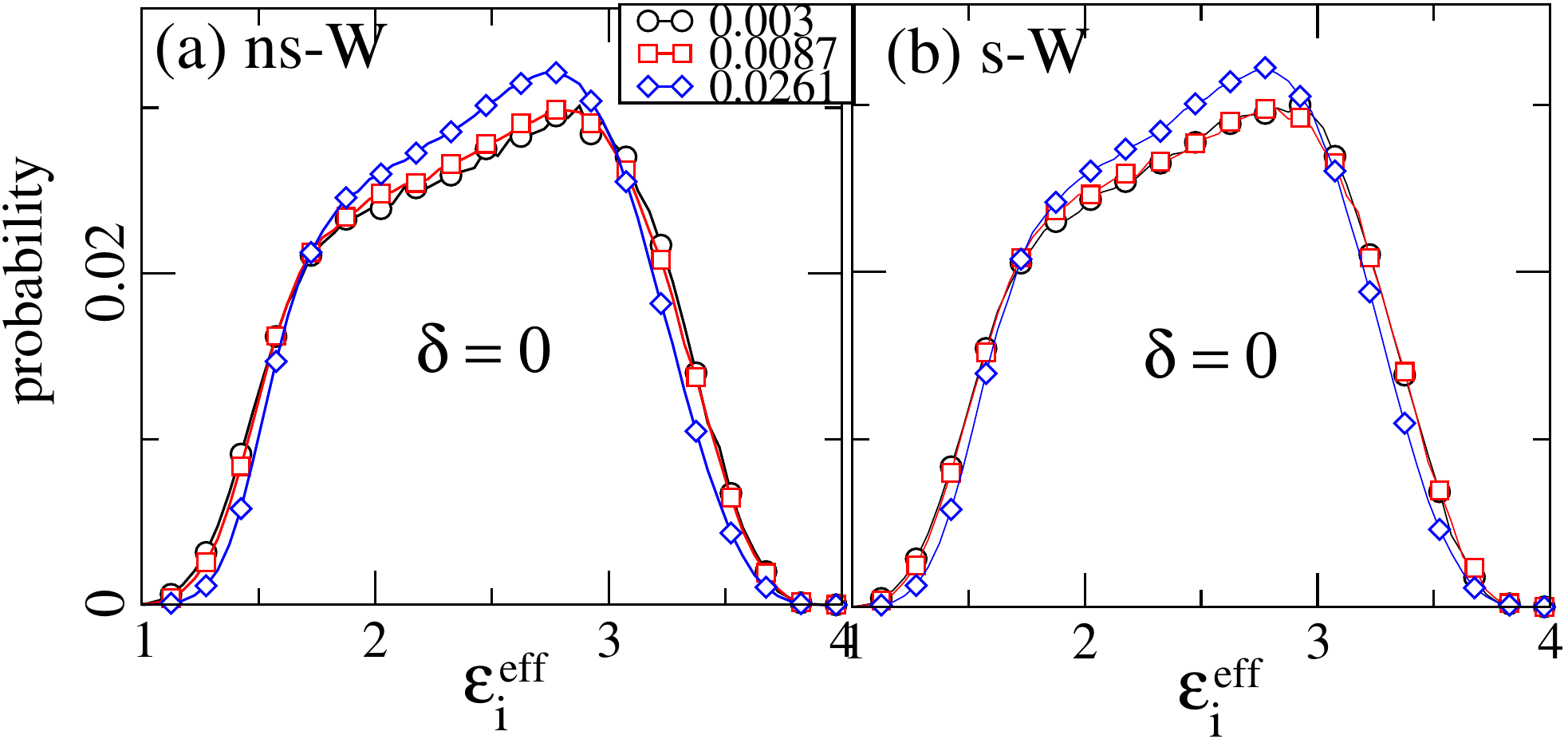}
	\begin{center}
		\caption{(Upper panel) Average effective interaction parameter $\epsilon_i^{\rm eff}$ as a function of normalized film thickness $y/D$ shown for non-selective wall (nsW) and selective wall (sW) types. (Lower panel) Distribution of $\epsilon_i^{\rm eff}$ at three different values of $\dot\gamma$ indicated in the figure at $\delta=-0.1,0$ for both nsW and sW.} 
		\label{fig: apd-eeff-compare1} 
	\end{center}
\end{figure}

\begin{figure}[ht]
	\begin{center}
	\includegraphics*[width=0.45\textwidth]{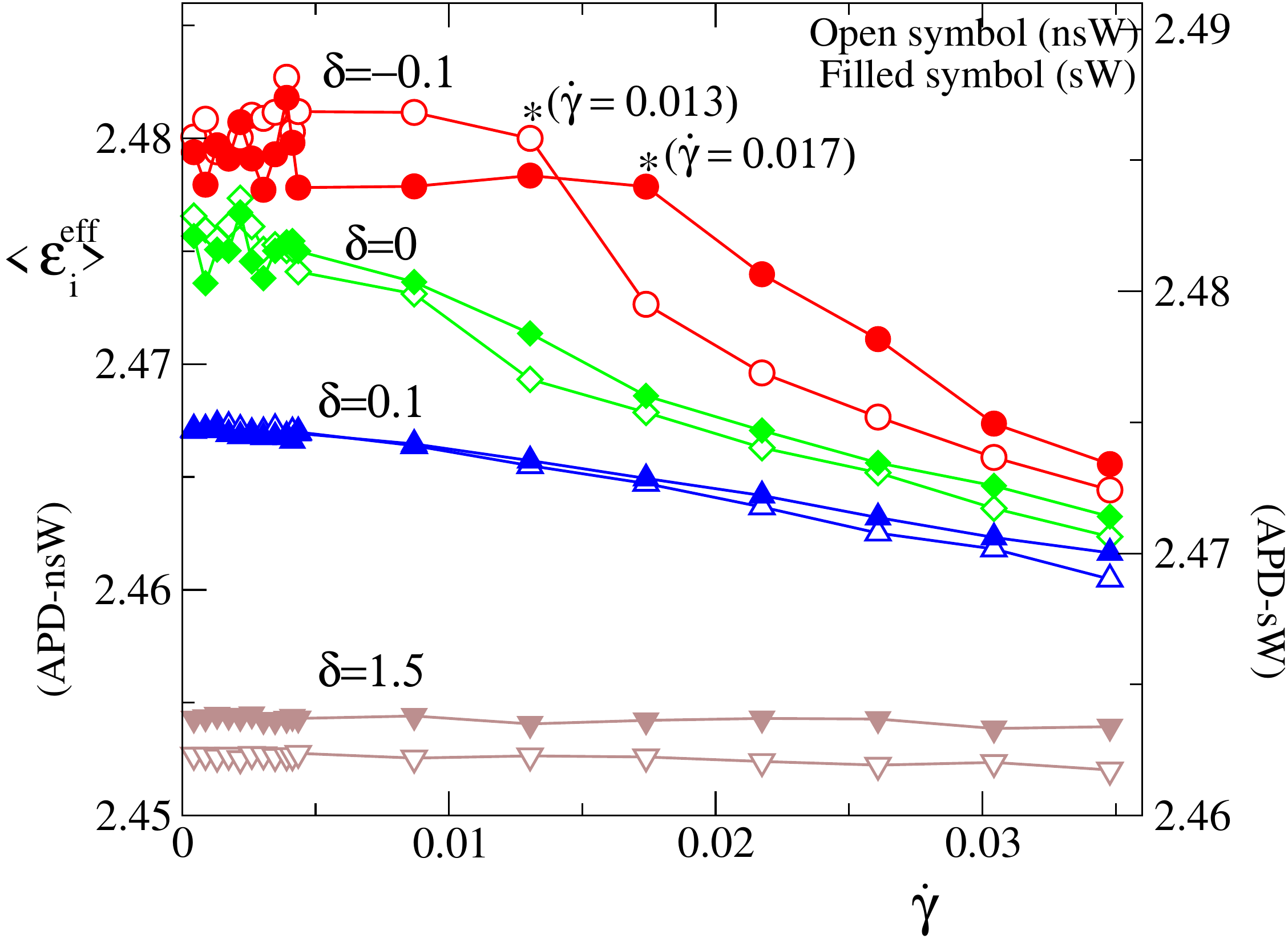}
	\caption{Two scale plot showing the variation of mean effective interaction parameter $\left< \epsilon_i^{\rm eff} \right>$ for APD system with nsW (left scale) and sW (right scale) obtained at different values of $\delta$ indicated in the figure. Open(close) symbols represent APD-nsW (APD-sW) system. Asterix marked points ($\dot\gamma\approx$ 0.013 and 0.017 for ns-W and s-W respectively) in the $\delta=-0.1$ curves indicates the shear rate above which $\left< \epsilon_i^{\rm eff} \right>$ decreases.} 
	\label{fig: apd-eeff-mean-compare} 
	\end{center}
\end{figure}

To gain further insight we proceed to understand the spatial ordering of particles' identity at different locations of the film characterized by laterally averaged $\epsilon_i^{\rm eff}$ profile and $\epsilon_i^{\rm eff}$ distributions as shown in figure~\ref{fig: apd-eeff-compare1}. We observe that the laterally averaged $\epsilon_i^{\rm eff}$ profile closely follows that of the corresponding density profile shown in figure~\ref{fig: apd-rho-compare}. 
For both APD systems, at a given temperature, when we increase shear rate the spatial distribution of particles as well as identity becomes homogeneous throught the film indicated by the flattening of the curve and thus the effect of substrate quality is insignificant at high shear rates. 
As shown, for $\delta \approx 0$ the difference in the profiles at small shear rates vanishes at very high shear rate. 
For $\delta > 0$, where the system is in fluid state, the profile is independent of $\dot \gamma$ (except for small differences close to the walls at small shear rates). On the other hand, $\epsilon_i^{\rm eff}$ distributions show that at high shear rates peak position is shifted to a lower value. 
The mean value $\left< \epsilon_i^{\rm eff} \right>$ at different shear rates is plotted in figure~\ref{fig: apd-eeff-mean-compare} for APD systems. Slightly below the transition temperature, $\left< \epsilon_i^{\rm eff} \right>$ roughly remains constant for small shear rates and decreases beyond a threshold value as indicated in the figure for $\delta \approx -0.1$. 
The decrease of $\left< \epsilon_i^{\rm eff} \right>$ with increasing shear rate is observed for all the temperatures considered. However, at high temperature (see $\delta \approx 1.5$ in the figure) the shear rate dependence of $\left< \epsilon_i^{\rm eff} \right>$ is rather weak. It is expected that the curves for relatively small $\delta$ (i.e., close to transition temperature $T^\ast$) will approach that of the high temperature of the respective systems at sufficiently high $\dot{\gamma}$ values, where the applied mechanical shear dominates over the interparticle interaction and the local identity ordering is completely destroyed. 
Between the two APD systems no qualitative difference is observed, but only a quantitative difference, i.e., for a given shear rate $\left< \epsilon_i^{\rm eff} \right>$ is relatively higher for APD-system with selective walls and hence a relatively large $\dot\gamma$ where $\left< \epsilon_i^{\rm eff} \right>$ drops for $\delta \approx -0.1$. The observed decrease in mean effective interation parameter with increasing shear rate is associated with the structural changes in the system and is consequently reflected in the macroscopic response of the system discussed in the following section.  

%

\section{Macroscopic response}  
\label{sec: macro-response}  
When mechanical shear is introduced to simple fluids, a wide variety of intersting effects such as shear induced ordering and phase transition, and non-Newtonian rheological behavior, i.e., shear thinning/thickening are observed.\cite{hess1980,hess1985,hess1994,hess1997,evans1984,evans2004} For the present APD fluid system also, the rheological behavior or the macroscopic response under shear is studied. In particular, we calculate shear viscosity $\eta_{xy}$ defined as
\begin{equation}
\eta_{xy}=\frac{\sigma_{xy}}{\dot\gamma}~,
\label{eq: shear-visco}
\end{equation}
with $\sigma_{xy}$ shear stress,  
in order to see whether there is any qualitative and quantitative differences among the systems. For Newtonian fluid shear viscosity is independent of shear rate, i.e.,
\begin{equation}
\eta_{xy}\sim\dot{\gamma}^0~.
\end{equation}
On the other hand, for non-Newtonian fluid  
\begin{equation}
\eta_{xy}\sim \dot{\gamma}^\alpha~,
\end{equation} 
in general, with $\alpha<0$ for shear-thinning and $\alpha>0$ for shear-thickening. 

\begin{figure}[ht]
	\begin{center}
	\includegraphics*[width=0.45\textwidth]{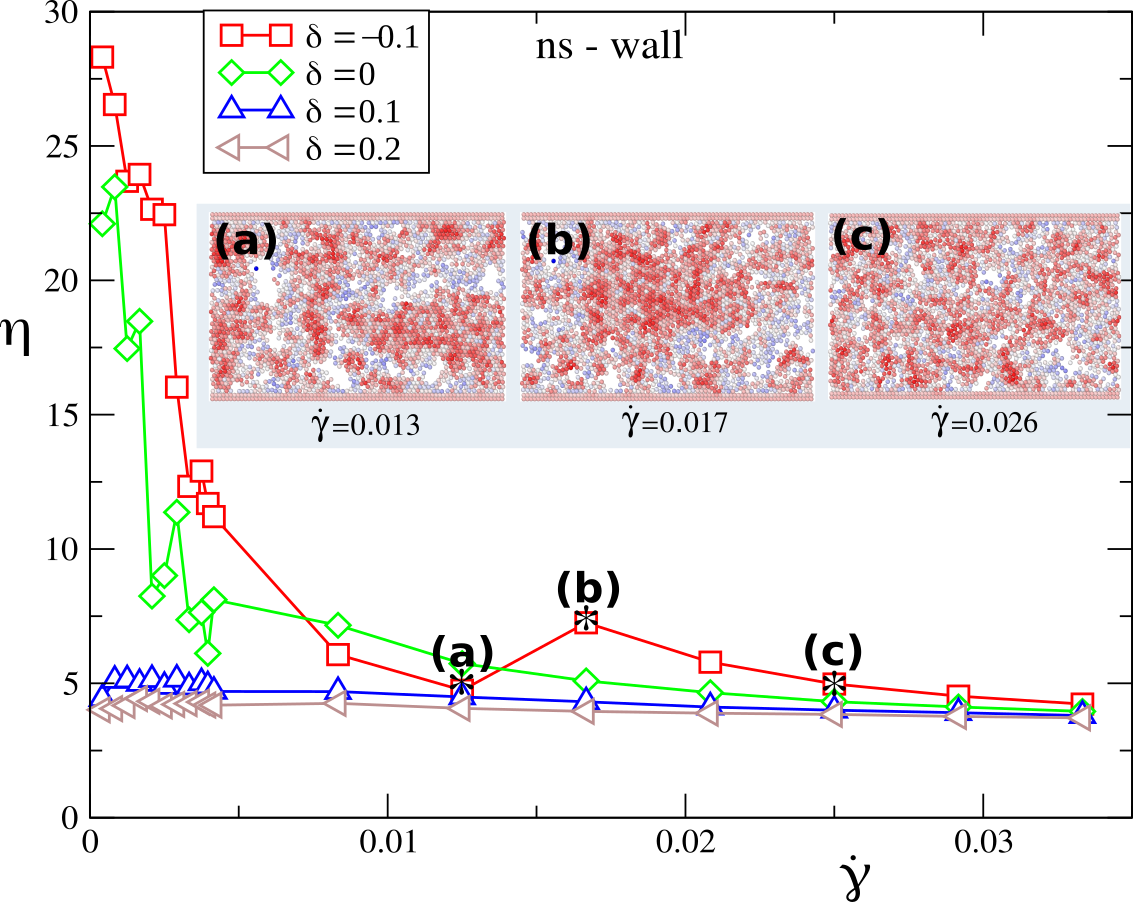}
	\caption{Shear viscosity $\eta$ against the imposed steady shear rate $\dot\gamma$ for APD system with non-selective wall at four different temperatures indicated in the figure. Subfigures maked (a)-(c) are the corresponding steady state configurations for the $\bigotimes$ marked points in the curve for $\delta=-0.1$.} 
	\label{fig: shear-visco-nsw}  
	\end{center}
\end{figure}

\begin{figure}[ht]
	\begin{center}
		\includegraphics*[width=0.45\textwidth]{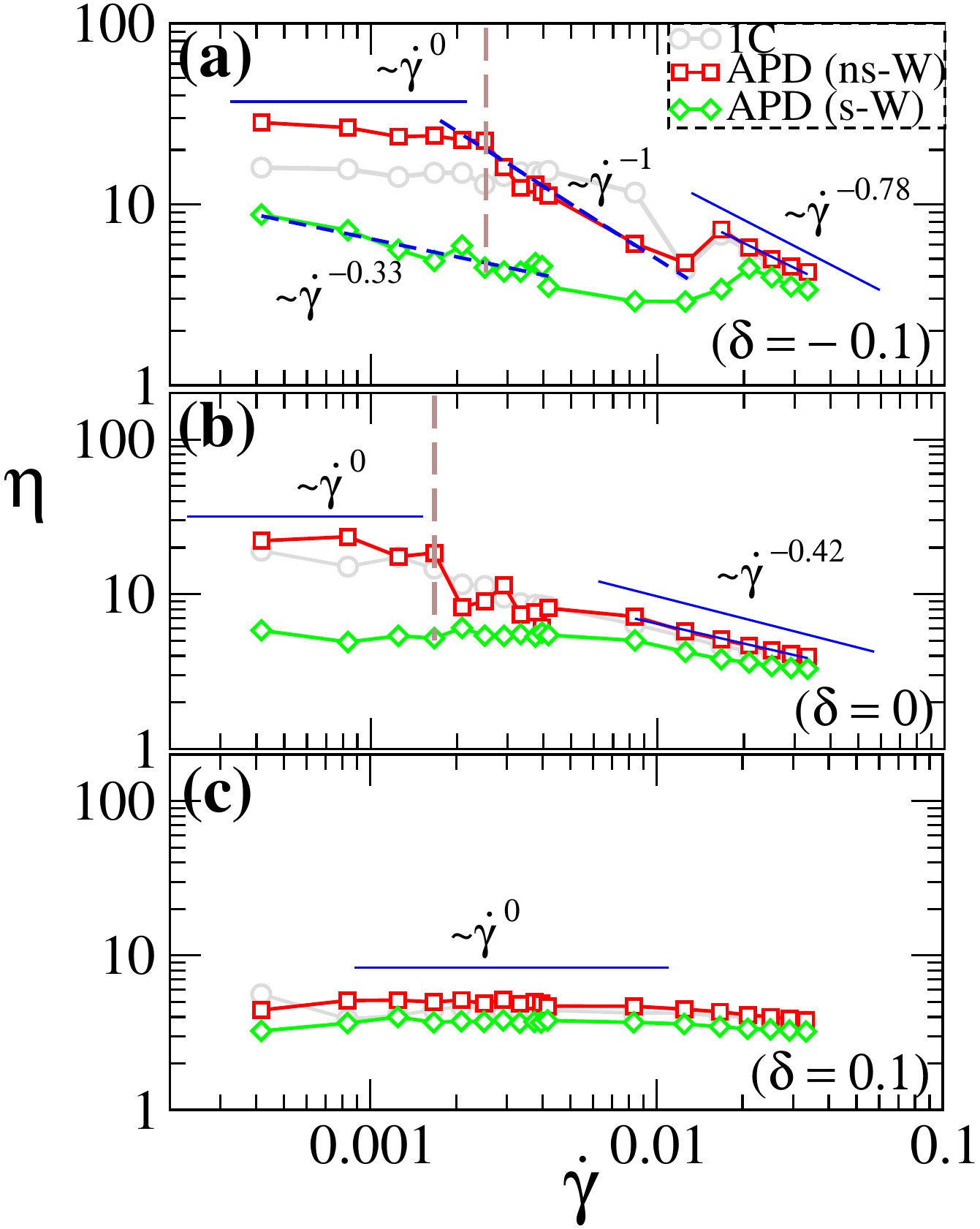}
		\caption{Comparison of shear viscosity $\eta_{xy}$ profiles among different systems indicated in the figure for (a) $\delta=-0.1$, (b) $\delta=0$, and (c) $\delta=0.1$. Vertical dashed line is guide to eye to indicate the onset shear thinning for APD system with ns-Walls.} 
		\label{fig: shear-visco-compare}  
	\end{center}
\end{figure} 
 
Typical shear viscosity profiles of the APD system (with non-selective substrates) close to transition temperature is shown in figure~\ref{fig: shear-visco-nsw}. For $T<T^*$ $(\delta=-0.1~\text{in the figure})$, it is observed that the shear viscosity $\eta_{xy}$ decreases with increasing $\dot\gamma$ (shear-thinning) which is followed by an increase in $\eta_{xy}$ on further increasing $\dot\gamma$ (shear-thickening) and finally a decrease again at high $\dot\gamma$ values (shear-thinning). 
The shear-thickening happens in a relatively short range of $\dot\gamma$, and the onset of shear-thickening shifts towards smaller $\dot\gamma$ values as $T\rightarrow T^*$, e.g., the onset is observed at $\dot\gamma\approx0.013$ for $\delta=-0.1$ as indicated in the figure. Further, it is interesting to note that the onset of shear-thickening corresponds to the decrease in average interaction parameter $\left< \epsilon_i^{\rm eff} \right>$, see figure~\ref{fig: apd-eeff-mean-compare} for $\delta=-0.1$ (nsW), and thus reflects the structural as well as local particle identity ordering changes with increasing shear rate. Steady-state configurations of the systems around the shear-thickening region is shown in figure~\ref{fig: shear-visco-nsw} as inset. It is evident from the configurations that before the onset of shear-thickening no major identity reorganization takes places and thus $\left< \epsilon_i^{\rm eff} \right>$ is roughly constant (e.g., see figure~\ref{fig: apd-eeff-mean-compare} for $\delta<0$). Comparison of subfigures -(a) and -(b) of figure~\ref{fig: shear-visco-nsw} clearly shows that a strong spatial and identity ordering happens during the shear-thickening, where we see particle clustering (formation of denser phase composed predominantly of high-$\epsilon_i^{\rm eff}$ particles). Similar behavior is observed during shear-thickening in colloidal systems.\cite{brown2014} The second shear-thinning corresponds to gradual melting and hence decrease in $\left< \epsilon_i^{\rm eff} \right>$ leading to a completely (identity) disordered fluid phase. When $\delta=0$ only shear-thinning is observed, and for $\delta>0$ shear-viscosity very weakly depends on the shear rate, i.e., $\eta_{xy}\sim \dot\gamma^0$. \\

In figure~\ref{fig: shear-visco-compare}, we show comparison among the systems at three different temperatures. Both shear-thinning and shear-thickening is observed for temperature slightly below $T^*$. As shown in figure~\ref{fig: shear-visco-compare}(a), at very small shear rates both APD-nsW and 1C systems show $\eta_{xy} \sim \dot{\gamma}^0$, and first shear-thinning ($\eta_{xy} \sim \dot{\gamma}^{-\alpha}$) sets in above a threshold value $\dot\gamma_c \approx 0.0025$ for APD-nsW (with exponent $\alpha \approx 1$) and $\dot\gamma \approx 0.004$ for 1C. On the other hand, for APD-sW system first shear-thinning is observed for shear rate upto $\dot\gamma \approx 0.01$ with exponent $\alpha\approx 0.33$. Comparing the exponent $\alpha$ it is clear that the shear-thinning is relatively strong for APD-nsW system. 
Upon increasing $\dot\gamma$ further shear-thickening is observed followed by second shear-thinning where the decrease of $\eta_{xy}$ with $\dot\gamma$ is first seen for APD-nsW and 1C systems followed by APD-sW system; however, all the systems have $\alpha\approx 0.78$. We note that as $\delta\rightarrow 0$ the onset of first shear-thinning shifts towards smaller value of $\dot\gamma$. At $\delta\approx 0$, the shear-thinning exponent $\alpha \approx 0.42$, see figure~\ref{fig: shear-visco-compare}(b). As $\delta$ becomes positive and gets larger (i.e. fluid phase further away from $T^*$) the exponent $\alpha\rightarrow 0$ and show Newtonian-like behavior for all systems.  

\section{Summary and discussion}
\label{sec: conclusion}

We have studied the effect of linear flow field on the spatial and particle identity ordering, density profiles, and macroscopic response of a multi-component fluid model. All the particles in the system are different in the sense that each particle has different interaction strength $\epsilon_i$ drawn from a uniform distribution. In equilibrium, this multi-component fluid model system exhibit local self-organization and clustering of particles according to their identity, i.e. a non-random state, leading to different dynamical properties for different particle types. When the system is subjected to steady flow field, depending on the shear rate, it displays various interesting features. We focus on temperature close to the solidification transition and investigated the system at various shear rates considering non-selective and selective substrate types. The results are briefly summarized below. 

The particle density profiles of the two APD systems is significantly different, e.g., in contrast to APD (ns-wall), the density profile for APD (s-Wall) has maxima at the center and minima (or particle depletion) near the substrates. As we increase the shear rate particle density becomes homogeneous and no difference is seen between the two APD-systems in the bulk, however close to the substrate significant difference in the density is observed enven at high shear rates. It is interesting to note that the density profiles of APD(ns-wall) and 1C systems are very similar.\\
Further study of radial distribution function reveal that this homogenization is related to the shear-induced melting, where higher order peaks (corresponding to hexagonal packing) in $g(r)$ disappears at high shear rates. 
Apart from structural changes, one of the main concern of this study is to understand how the neighborhood identity ordering is affected due to shear. As detailed in section~\ref{sec: identity-ordering}, the neighborhood identity ordering is characterized through $\epsilon_i^{\rm eff}$. The laterally averaged $\epsilon_i^{\rm eff}$ profiles closely follow that of the corresponding density profiles, where the value of $\epsilon_i^{\rm eff}$ is higher in higher density regions, see figure~\ref{fig: apd-eeff-compare1}. At high shear rates, in consistent with the homogenization observed, the profile is flat and $\epsilon_i^{\rm eff}\rightarrow 2.5$ in all the regions (except for small deviations near the substrates). Furthermore, the $\epsilon_i^{\rm eff}$ distribution show that the mean of the distribution is shifted to lower values with increasing shear rates and hence $\left<\epsilon_i^{\rm eff}\right>$ also decreases. We found only quantitative difference between the systems with selective and non-selective substrates, i.e., $\left<\epsilon_i^{\rm eff}\right>$ for system with selective-walls is consistently higher in the range of $\dot\gamma$ considered. It is clear from this observation that the local particle identity ordering is affected by the quality of substrates and it is reflected in the global average quantity (i.e., selective substrate type leads to higher value of $\left<\epsilon_i^{\rm eff}\right>$). 

Finally, we looked at the behavior of shear viscosity, $\eta_{xy}$, for both systems. Below the transiton temperature both energy polydispersed and 1C systems show shear-thinning as well as shear-thickening regimes, and shear-thickening is more pronounced with temperature farther away from the transition point. Shear-thickening is also associated with transient increase of $\left<\epsilon_i^{\rm eff}\right>$. Shear-thickening is followed by shear-thinning again leading to shear-induced melting at high shear rates. Exponent characterising shear-thinning has braod distribution and depends on temperature. It is important to point out that APD fluid with selective (non-selective) substrate has the shear viscosity below (or above depending on $T$) that of 1C fluid indicating that by tuning substrate or wall properties one can tune shear viscosity of this multi-component fluid. In conclusion, we present a comprehensive study on the energy polydispersed fluid models in the presence of flow field and we believe that this model, apart from biological systems, is also very relevant to industrial mixing of materials containing many energetically different component.

\begin{acknowledgements}
LS acknowledge fruitful discussion with Yitzhak Rabin. 
\end{acknowledgements}


\end{document}